\documentclass[10pt]{article}

\usepackage{lineno,hyperref}
\usepackage{graphicx}
\usepackage{color}
\usepackage{pgf}
\usepackage{amsmath}
\usepackage{amssymb}
\usepackage{subfigure}
\usepackage{multirow}
\usepackage[latin9]{inputenc}
\usepackage{hyperref}
\usepackage{breakurl}
\usepackage{pgf}
\usepackage{subfigure}
\usepackage{multirow}
\usepackage{bbm}
\usepackage{booktabs} 
\usepackage{newtxtext}
\usepackage{float}
\usepackage{soul}

\usepackage[labelfont=bf]{caption}

\usepackage{authblk}

\author[1]{Ruijin Cang}
\author[2]{Yaopengxiao Xu}
\author[2]{Shaohua Chen}
\author[1]{Yongming Liu}
\author[2]{Yang Jiao}
\author[1]{Max Yi Ren}
\affil[1]{Department of Mechanical Engineering, Arizona State University, Tempe}
\affil[1]{Department of Materials Science and Engineering, Arizona State University, Tempe}

\iffalse
        \newcommand{\cutsectionup}{\vspace*{-0.1in}}

        \newcommand{\cutsubsectionup}{\vspace*{-0.09in}}

        \newcommand{\cutequationup}{\vspace*{-0.12in}}
        \newcommand{\cutequationdown}{\vspace*{-0.12in}}

\else 
        \newcommand{\cutsectionup}{}

        \newcommand{\cutsubsectionup}{}

        \newcommand{\cutequationup}{}
        \newcommand{\cutequationdown}{}

\fi

\iftrue 

\newcommand{\todo}[1]{{\textcolor{red}{[[TODO: {#1}]]}}}
\newcommand{\commenttext}[1]{{\textcolor{red}{[[{#1}]]}}}
\newcommand{\commentfoot}[1]{\footnote{\textcolor{red}{\emph{Comment: #1}}}}
\newcommand{\topic}[1]{}
\else
\newcommand{\todo}[1]{}
\newcommand{\commenttext}[1]{}
\newcommand{\commentfoot}[1]{}
\newcommand{\topic}[1]{}
\fi

\begin{document}

\title{Microstructure Representation and Reconstruction of Heterogeneous Materials via Deep Belief Network for Computational Material Design\footnote{The paper is accepted by JMD}}

\maketitle

\begin{abstract}
Integrated Computational Materials Engineering (ICME) aims to accelerate optimal design of complex material systems by integrating material science and design automation. For tractable ICME, it is required that (1) a structural feature space be identified to allow reconstruction of new designs, and (2) the reconstruction process be property-preserving. The majority of existing structural presentation schemes rely on the designer's understanding of specific material systems to identify geometric and statistical features, which could be biased and insufficient for reconstructing physically meaningful microstructures of complex material systems. In this paper, we develop a feature learning mechanism based on convolutional deep belief network to automate a two-way conversion between microstructures and their lower-dimensional feature representations,and to achieve a 1000-fold dimension reduction from the microstructure space. The proposed model is applied to a wide spectrum of heterogeneous material systems with distinct microstructural features including Ti-6Al-4V alloy, Pb63-Sn37 alloy, Fontainebleau sandstone, and Spherical colloids, to produce material reconstructions that are close to the original samples with respect to 2-point correlation functions and mean critical fracture strength. This capability is not achieved by existing synthesis methods that rely on the Markovian assumption of material microstructures.

\end{abstract}



\section{Introduction}
\label{sec:intro}
Integrated Computational Materials Engineering (ICME) has been promoted by the Material Genome Initiative to fundamentally change the strategies for developing and manufacturing advanced material systems to meet the urging demands in energy, health and national security. Pioneering ICME works have demonstrated successes in designing a variety of material systems, including functional polymers~\cite{sharma2014rational,baldwin2015rational,ma2015rational}, alloys and ceramics~\cite{kalidindi2015materials,kaczmarowski2015genetic,kirklin2016high}, and polymer-matrix composites~\cite{xu2014descriptor,xu2015machine}, to name a few. Nonetheless, existing ICME methods lack scalability and have limited applications to multiscale and high-resolution microstructures. To this end, this paper proposes a computational tool for extraction and reconstruction of microstructure features on multiple length scales. The developed methodology from this paper could lead to a better definition of the design space and enable computational design of complex material systems. In the remainder of this section, we introduce the mathematical formulation of the material microstructure design problem and review contemporary feature extraction techniques. Specifications of the proposed model and demonstrations of its performance can be found in Section \ref{sec:model}. The advantages and limitations of the proposed algorithm will be discussed in Section \ref{sec:discussion}, followed by conclusions in Section \ref{sec:con}.

\subsection{Material design as an optimization problem}
We start by providing an overview of ICME to justify the necessity of design at the structure level. Computational material design can be mathematically formulated as follows:
A material microstructure is represented as a 2D or 3D image ${\bf z} \in \mathcal{Z}$, where $\mathcal{Z}$ is the image space. Each element of ${\bf z}$ defines the composition or phase at its location. The image is called {\it feasible} when some processing setting $\boldsymbol{\boldsymbol{\theta}} \in \Theta$ exists, such that $z = p(\boldsymbol{\theta})$, with $p(\cdot)$ being the process-structure mapping, e.g., a physics-based simulation, and $\Theta$ the set of all available processing settings. Consider a material property of interest, e.g., critical fracture force of alloy, derived from a structure-property mapping $f(\cdot)$: $y = f({\bf z})$. Finding the optimal material property $y^*$ through an all-in-one approach, i.e., $\boldsymbol{\theta}^* = arg\min_{\boldsymbol{\theta} \in \Theta} \quad y = f(p(\boldsymbol{\theta}))$, may not be favorable for the following reasons: (1) Finding the optimal $\boldsymbol{\theta}^*$ can be computationally intractable if the problem is combinatorial, e.g., a combination of multiple processing techniques is needed. (2) The mapping $p(\boldsymbol{\theta})$ is often stochastic, i.e., a certain processing setting can lead to a set of microstructure images. One solution is to perform a nested optimization by first finding a target microstructure ${\bf z}^*$ that optimizes $y$, then searching for a processing setting $\boldsymbol{\theta}^*$ to match ${\bf z}^*$. This approach can be mathematically formulated as the following structure design problem constrained by processing feasibility: 
\cutequationup
\begin{equation}
\min_{\text{feasible } {\bf z}} \quad y = f({\bf z}).
\label{eq:structureopt}
\cutequationdown
\end{equation}

\cutsubsectionup
\subsection{Scientific challenges in optimal material structure design}
\label{sec:challenge}
For complex material systems, solving Eq. \eqref{eq:structureopt} will encounter the following issues: The high sensitivity of material properties on microstructure details requires high-resolution microstructure images leading to high-dimensional $\mathcal{Z}$; and the nonlinear processing-structure mapping makes the feasible domain within $\mathcal{Z}$ costly to be characterized. These issues can be alleviated if the dimensionality of the design space can be reduced. Dimension reduction is possible due to the existence of common structural patterns in the given material system, indicating that there exists a feature representation that concisely encodes the raw material images. Further, since the structural patterns are often related to processing feasibility and material properties, the encoding may lead to easier-to-construct predictive models for the processing-structure-property mappings. Advantages of feature representations were confirmed by recent studies for homogeneous nanoparticle composite design tasks~\cite{xu2014descriptor,xu2015machine}. However, current practices rely on the designer's understanding of specific material systems to identify geometric (e.g., particle size and orientation) and statistical descriptors (e.g., correlation functions) that are likely to explain variances in material properties. Generating such quantitative features for complex material systems could be difficult for human designers, or, the manually defined feature set would be insufficient for reconstruction of physically meaningful microstructures. Feature extraction methods through the use of deep convolutional networks could potentially address this issue. Informally, this line of approaches use sampled microstructures to learn a mapping ${\bf x}=\Phi({\bf z})$ and its inverse ${\bf z}= \Phi^{-1}({\bf x})$ (the unsupervised learning step), where ${\bf x}$ is a learned feature representation of ${\bf z}$ from a feature space $\mathcal{X}$, before mapping ${\bf x}$ to the performance $y$ (the supervised learning step). 

A unique challenge in developing a feature extraction mechanism for (material) design is the demanding quality of decoding (or called reconstruction), i.e., the reconstruction $\Phi(\Phi^{-1}({\bf z}))$ shall ``match'' ${\bf z}$, so that any design searched in $\mathcal{X}$ can be correctly validated through a physics-based simulation or an experiment. In Sec.~\ref{sec:validation}, we examine both the visual appearances, the 2-point correlation functions, and the mean property values of $\Phi(\Phi^{-1}({\bf z}))$ and ${\bf z}$. Visually meaningful decoding is yet challenging for existing deep networks, which tends to neglect local details while preserving key features during the  reconstruction~\cite{dosovitskiy2015inverting,mahendran2015understanding,nguyen2015deep}. To illustrate, photos can be encoded as text labels with a state-of-the-art deep network (e.g., \cite{krizhevsky2012imagenet,simonyan2014very}) but image restoration from text labels is hard to be precise~\cite{nguyen2015deep,yan2015attribute2image}. Existing studies have achieved limited success at this challenge by imposing structures onto the image space, e.g., constraining the smoothness of the reconstructed image~\cite{dosovitskiy2015inverting,mahendran2015understanding,yan2015attribute2image}. This study first investigates a deep network model and postprocessing steps to enable random reconstruction of the complex microstructure of Ti-6Al-4V alloy samples. We then show that the proposed model can be applied to other material systems, including Pb-Sn (lead-tin) alloy, Pore structure of Fontainebleau sandstone, and 2D suspension of spherical colloids.

\cutsubsectionup
\subsection{Contributions and limitations}
To the authors knowledge, this is the first study that develops a deep network model for both feature extraction and reconstruction of complex material systems. In addition, we show that our method preserves material properties statistically through reconstruction. These properties include the 2-point correlation function and the critical fracture strength.
Two major limitations still exist: First, while computationally inexpensive, the current reconstruction process is by nature non-scalable. A hybrid approach that combines the proposed feature extraction network and a stochastic material synthesis model will be discussed. Second, the mechanical properties of individual samples are not preserved, demanding higher reconstruction quality and better preservation of property-related microstructure features through the network model.

\cutsectionup
\section{Related Work}
\label{sec:lit}
Design at the microstructure level by changing the composition, phases and morphologies could lead to the discovery of unprecedented material systems with advanced performance~\cite{McDowell2008}. The challenges, however, exist in the modeling of interactions among multiple material constituents to predict the properties, and the automation of material design at microstructure level. In the following, we review existing studies that concern the latter challenge, and then introduce background knowledge for the proposed approach based on Convolutional Deep Belief Network (CDBN). 

\cutsubsectionup
\subsection{Microstructure parametrization and reconstruction}
Conventional material design relies on choosing different material compositions from material databases~\cite{Broderick2008}. For example, carbon-fibre reinforced epoxy are applied to replace the wooden oars for better performance~\cite{ashby1993materials}. This composition-based approach is often limited and cannot be applied in the design of complex materials systems due to the existence of features other than constituents that govern material properties. Such features include morphology of microstructure, i.e., the spatial arrangement of local microstructural features~\cite{Karasek1996}, the heterogeneity in microstructure~\cite{xu2014descriptor} and others. A quantitative representation of material systems is thus needed. 

\begin{figure}[t]
\centering
\graphicspath{ {image/} }
\includegraphics[width=\linewidth]{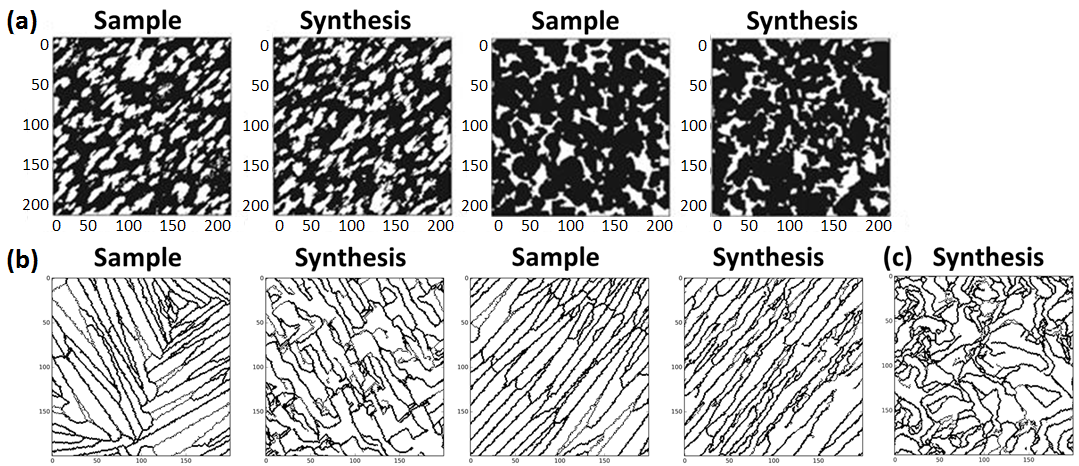}
\caption{\textbf{(a) Samples and random synthesis results of material systems that are assumed to be Markovian~\cite{bostanabad2016stochastic}. (b-c) Random synthesis of the Ti-6Al-4V alloy microstructure following the synthesis algorithm from \cite{bostanabad2016stochastic}. The synthesis is based on (b) a single sample and (c) 100 samples from Fig. \ref{fig:mat1}. Image courtesy of Dr. Ramin Bostanabad.}}
\label{fig:Wei}
\end{figure}

A review from \cite{xu2014descriptor} summarizes three categories of microstructure representations: (1) physical descriptors (e.g., composition descriptors such as volume fraction, dispersion descriptors~\cite{rollett2007three, borbely2004three, tewari2004nearest, pytz2004microstructure, steinzig1999probability,scalon2003spatial} such as average distance between fillers, and geometry descriptors~\cite{rollett2007three, steinzig1999probability, torquato2013random, sundararaghavan2005classification, basanta2005using, holotescu2011prediction, klaysom2011effects, gruber2010misorientation} such as the size of fillers), (2) 2-point or $N$-point correlation functions~\cite{liu2013computational,borbely2004three, torquato2013random, sundararaghavan2005classification, basanta2005using}, and (3) random fields~\cite{torquato2013random,quiblier1984new, jiang2013efficient, grigoriu2003random}. 
Accordingly, a variety of material reconstruction schemes have been devised for different microstructure representations. Examples include the Gaussian random field method\cite{roberts1997statistical}, stochastic optimization method \cite{yeong1998reconstructing,jiao2008modeling,jiao2009superior,karsanina2015universal}, gradient-based method \cite{fullwood2008gradient}, phase-retrieval method\cite{fullwood2008microstructure}, iterative methods based on multi-point statistics\cite{okabe2005pore, hajizadeh2011multiple}, Bayesian Network Method\cite{matthews2016hierarchical} and cross-correlation functions \cite{tahmasebi2013cross,tahmasebi2015reconstruction}, and image synthesis method \cite{sundararaghavan2005classification, liu2015random}. Recently, a new reconstruction method is proved to be efficient at synthesizing {\it Markovian}
microstructures~\cite{bostanabad2016stochastic} where the probability distribution of the material composition at each pixel (or voxel) is determined by its {\it local} surroundings and the conditional probability model can be applied {\it homogeneously} across a microstructure sample, see Fig.~\ref{fig:Wei}a for examples. However, the Markovian assumption may not hold for complex material systems. In fact, the algorithm from \cite{bostanabad2016stochastic} produces less plausible synthesis results when trained on Ti-6Al-4V alloy samples, see Fig.~\ref{fig:Wei}b and c, especially when high variances in grain directions and sizes exist.


\cutsubsectionup
\subsection{Convolutional deep network}
\begin{figure}[htp]
\centering
\graphicspath{ {image/} }
\includegraphics[width=\linewidth]{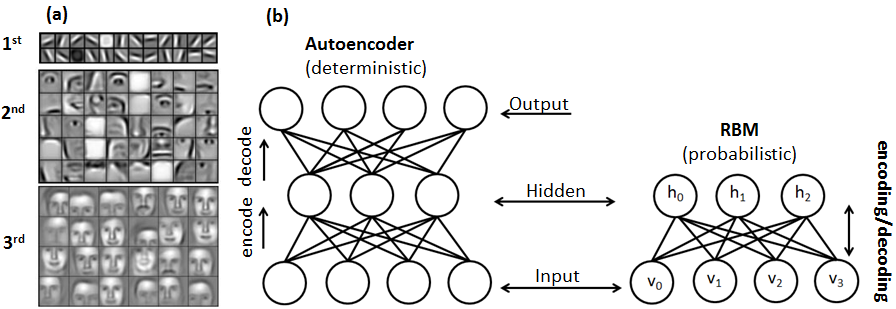}
\caption{\textbf{(a) Three layers of human face features extracted at increasing length scales~\cite{lee2009convolutional}. (b) A schematic comparison between an RBM and an autoencoder. $a_i$ ($b_j$) and $v_i$ ($h_j$) are the bias and state values of visible (hidden) layer, respectively. $W_{ij}$ indicates the weights between the visible and hidden layers.}}
\label{fig:ACandRBM}
\end{figure}

Convolutional networks have long been used for feature extraction, with applications to object and voice recognition~\cite{hinton2012deep} and detection~\cite{NIPS2012_4824}, reinforcement learning~\cite{mnih2013playing,schmidhuber2015deep,levine2015end}, analogy making~\cite{reed2015deep} and many others. By stacking multiple layers, a convolutional deep network can learn from input training data multi-scale features that contribute to the explanation of the corresponding outputs, without necessarily being guided by the outputs. For example, using a large set of human face images, a network can extract various face elements, including basic edges from the $1$st layer, eyes and noses from the $2$nd, and partial faces from the $3$rd, see Fig.~\ref{fig:ACandRBM}a. A forward pass of an image through the trained network essentially checks the existence of these features in the image, and thus using the output of the network for prediction (e.g., face recognition) is more effective than using raw image pixel values~\cite{lee2009convolutional}. While a deep network may involve a large number of model parameters, the training is made tractable by using stochastic gradient descent~\cite{bousquet2008tradeoffs} along with back-propagation (for deterministic models)~\cite{rumelhart1988learning} and contrastive divergence (for probabilistic models such as Restricted Boltzmann Machines (RBM))~\cite{hinton2002training}. 

The proposed Convolutional Deep Belief Network is one type of deep network that consists of stacked RBM layers.
By the nature of RBM, the network is generative, meaning that it can perform both dimension reduction to map an input microstructure to a low-dimensional feature space, and reconstruction of the microstructure for a given sample in the feature space. It should be noted that other models could also be developed for reconstruction purpose, including Autoencoder~\cite{bengio2009learning} (see Fig.~\ref{fig:ACandRBM}b for an illustration and ~\cite{yumer2015procedural} for an example), Variational Autoencoder (VAE)~\cite{kingma2013auto,yan2015attribute2image}, and the Generative Adversarial Network (GAN)~\cite{goodfellow2014generative}. 

\cutsubsectionup
\subsection{Restricted Boltzmann Machine}
\label{sec:rbm}
A fully-connected RBM layer consists of visible (input) and hidden (output) nodes, see Fig.~\ref{fig:ACandRBM}b. With weighted edges connecting these two sets of nodes, it is a complete bipartite graph. The set of edges connecting from all visible nodes to one hidden node acts as a convolution filter applied to the input image. The output of the convolution is transformed by a sigmoid function, and treated as the {\it activation} of the corresponding hidden nodes. The {\it state} of each hidden node is a binary number drawn from a Bernoulli distribution parameterized by the activation. A Convolutional RBM (or CRBM) is a special type of RBM where all image batches in the input layer (of the size of the filters) share $K$ filters, so that the hidden layer forms a binary image with $K$ channels, each produced by convolving the input image with a filter. Thus, a CRBM only has a subset of the visible nodes connecting to each of the hidden nodes (see Fig.~\ref{fig:CDBN3}a). 
In the rest of the paper, the term ``filter'' and ``feature'' are used interchangeably.

Learning an RBM (and CRBM) model through an set of input images takes the following procedure. Notations follow Fig.~\ref{fig:ACandRBM}. Consider a model with $n$ visible and $m$ hidden nodes, and let $v_i$ and $h_j$ be the states of the $i$th visible and the $j$th hidden nodes, respectively. When visible units are binary-valued, the total energy of the layer is defined as:
\cutequationup
\begin{equation}
     E({\bf v},{\bf h};\boldsymbol{\theta})=-\sum\limits_{i=1}^n a_iv_i - \sum\limits_{j=1}^m b_jh_j - \sum\limits_{i=1}^n\sum\limits_{j=1}^mv_iW_{ij}h_j.
\cutequationdown
\label{eq:totalenergy1}
\end{equation}
In the above equation, $\boldsymbol{\theta}:=\{{\bf W},{\bf a},{\bf b}\}$ is a set of model parameters to be estimated, within which ${\bf W}$ is a matrix of network weights, and ${\bf a}$ and ${\bf b}$ are the biases in the visible and hidden nodes, respectively. The joint probability of visible and hidden states is defined based on this energy function as:
\cutequationup
\begin{equation}
\text{Pr}({\bf v},{\bf h}|\boldsymbol{\theta})=\frac{\exp({-E({\bf v},{\bf h}|\boldsymbol{\theta})})}{Z(\boldsymbol{\theta})},
\cutequationdown
\label{eq:jointrbm}
\end{equation}
where $Z(\boldsymbol{\theta})=\sum\limits_{{\bf v},{\bf h}}\exp({-E({\bf v},{\bf h}|\boldsymbol{\theta})})$ is a partition function to make $\text{Pr}({\bf v},{\bf h}|\boldsymbol{\theta})$ a valid probability mass function. The marginal probability of the visible state is: 
\cutequationup
\begin{equation}
\text{Pr}({\bf v}|\boldsymbol{\theta})=\frac{1}{Z(\boldsymbol{\theta})}\sum{e^{-E({\bf v},{\bf h}|\boldsymbol{\theta})}}.
\cutequationdown
\label{eq:marginalrbm}
\end{equation}
Thus, the log-likelihood of the model parameters $\boldsymbol{\theta}$, given training data at the visible layer, can be written as $\phi(\boldsymbol{\theta};{\bf v})=\log(\text{Pr}({\bf v}|\boldsymbol{\theta}))=\phi^+ - \phi^-$,
where $\phi^+=\log\sum\nolimits_{{\bf h}}\exp({-E({\bf v},{\bf h})})$ and $\phi^-=\log{Z}=\log\sum\nolimits_{{\bf v},{\bf h}}\exp({-E({\bf v},{\bf h})})$ are called the positive and negative parts. Finding the maximum likelihood estimator of $\boldsymbol{\theta}$ requires applying stochastic gradient descend to $\phi(\boldsymbol{\theta};{\bf v})$. The gradient of the positive part can be simply calculated as $\frac{\partial \phi^+}{\partial W_{ij}}=v_i \text{Pr}(h_j = 1|v)$, while the negative gradient $\frac{\partial \phi^-}{\partial W_{ij}}=\text{Pr}(v_i = 1,h_j = 1)$ is usually computationally expensive due to the calculation of the partition function. This issue, however, can be resolved by using contrastive divergence~\cite{hinton2002training} to provide an asymptotic approximation of $\frac{\partial \phi^-}{\partial W_{ij}}$ by iteratively sampling from the conditional probabilities $\text{Pr}(v_i=1|{\bf h})=\text{sigmoid}\left(\sum_{j}W_{ij}h_j+a_i\right)$ and $\text{Pr}(h_j=1|{\bf v})=\text{sigmoid}\left(\sum_{i}W_{ij}v_i+b_j\right)$.

\cutsubsectionup
\subsection{Probabilistic max-pooling}
\begin{figure}[htp]
\centering
\graphicspath{ {image/} }
\includegraphics[width=\linewidth]{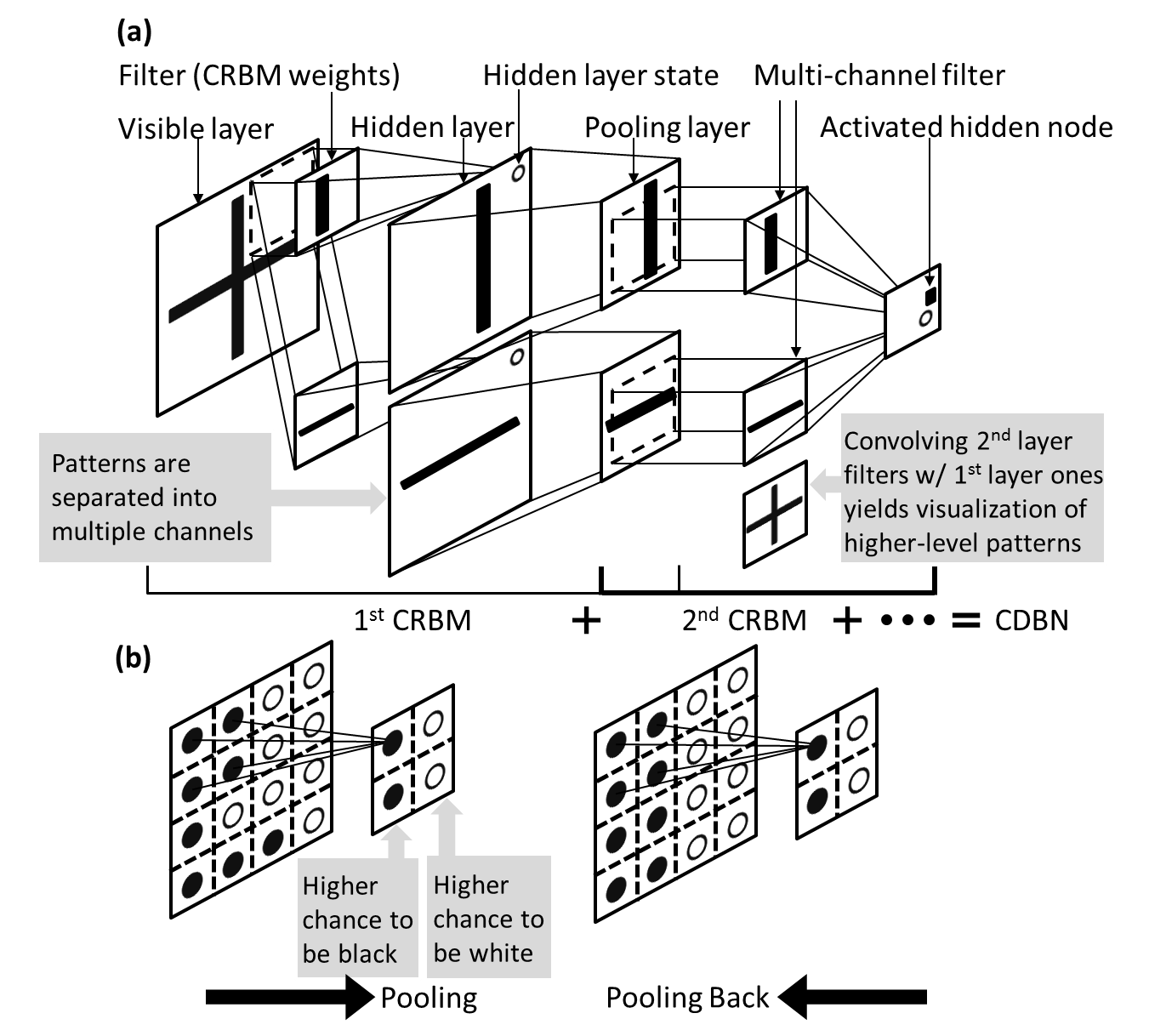}
\caption{\textbf{(a) CRBM and pooling layers (b) Forward and backward probabilistic max-pooling procedures with $2\times2$ blocks}}
\label{fig:CDBN3}
\end{figure}

In a deep network, a max-pooling layer is often inserted after a convolution layer to reduce the number of hidden nodes. This allows fewer network weights to be computed in the follow-up layers and also allow features of larger length scales to be extracted. Specific to CDBN, a probabilistic max-pooling layer segments the output image of a hidden RBM layer into blocks ($2\times2$ in this study) and assigns the associated pooling node a binary value $\alpha$ by drawing from a Bernoulli distribution defined by the activations of the hidden nodes. The activations for the four hidden nodes are denoted by $I_i$ for $i=1,2,3,4$. The probability for the pooling node to be activated is $\text{Pr}(\alpha=1) = 1-(1+\sum_{i=1}^{4}I_i)^{-1}$. For backward pooling during the reconstruction, we assign ones to all four nodes in the block if $\alpha=1$ or otherwise zeros. Fig.~\ref{fig:CDBN3}b summarizes the pooling procedure.

\cutsectionup
\section{Proposed CDBN for feature extraction and material reconstruction}
\label{sec:model}
The proposed network has three CRBM layers and two fully-connected RBM layers. Each of the first two CRBM layers is followed by a $2\times2$ probabilistic max-pooling layer. The convolutional filters for the first three layers are set respectively as follows: $6\times6$, $9\times9$, and $9\times9$ in filter size, $1$, $24$ and $40$ in the number of channels, and $24$, $40$ and $288$ in the number of filters. The size of the five hidden layers are correspondingly $200\times200\times1$, $97\times97\times24$, $44\times44\times40$, $36\times36\times288$ and $30\times1$. Notice that the image size shrinks along the depth of the network due to the convolution operation. For example, convolving an image of $200\times200$ with the filter of $6\times6$ will lead to an output of $(200-6+1)\times(200-6+1)$. This size will further be divided by $4$ after the max-pooling process, which makes the dimension $97\times97$. In addition, the number of filters corresponds to the number of channels of the hidden layer. Table~\ref{table:Parameter} summarizes the model and training parameters. One typical measure to prevent the learning of trivial filters is to set a target sparsity for the hidden nodes, so that on average a filter can only activate a limited number of hidden nodes and thus represents a specific image pattern. For the first four layers, a target sparsity of $0.1$ is imposed on the hidden layer activations. The $p_{\lambda}$ values specify the weights on the gradient of the sparsity penalty during the training. Both target sparsity and $p_{\lambda}$ values are manually tuned so that the training of the network can achieve low reconstruction error. During network training, we set the learning rate to be $0.005$, initial momentum to be $0.1$, final momentum to be $0.9$, and use one iteration of contrastive divergence for gradient approximation. The network weights are initialized  by drawing from independent normal distributions with a standard deviation of $0.02$. In addition, with $2,000$ epochs training for the first four training layer and $20,000$ for the last layer is applied in our cases. On a workstation withIntel Xeon E5-1620 v3 @3.5GHz CPU, GeForce GTX Titan GPU and $64$GB Memory, the training takes $14134$ seconds with $100$ input images of size $200\times200$ for each and $1363130$ network parameters.

\begin{table}[H]
\caption{\textbf{Network and algorithmic parameters}}
\centering
\label{table:Parameter}
\begin{tabular}{|l|l|l|l|l|l|l|}
\hline
Layer    & \#filter & filter size & sparsity & $p_{\lambda}$ & rotation \\ \hline
$1^{st}$ & $2$             & $6\times 6$          & 0.1      & 10     & 12       \\ \hline
$2^{nd}$ & $40$            & $9\times 9$           & 0.1      & 10     & N/A      \\ \hline
$3^{rd}$ & $288$           & $9\times 9$           & 0.1      & 10     & N/A      \\ \hline
$4^{th}$ & $1000$          & $36\times 36$         & 0.1      & 10     & N/A      \\ \hline
$5^{th}$ & $30$            & $1000\times 1$           & 0        & 0      & N/A      \\ \hline
\end{tabular}%
\end{table}

\cutsubsectionup
\subsection{Orientation-invariant filters}
\label{sec:orientation}
One common trait of microstructure images is the existence of low-level features that are similar under linear transformations. For example, grain boundaries of alloy samples in Fig.~\ref{fig:Wei}b are only different by orientations (of elongated grains). Therefore filters of the first CRBM layers can be parametrized by linear transformations, so that training of a small set of filters will be sufficient to capture all features, thus significantly accelerate the training process. Such as the material system of Ti64 Fig~\ref{fig:filter}, which is composed by lines with different orientation visually, so the basic features for these kinds of material system should be edges with various orientations. In our proposed model, $K$ = 2 rotation-invariant filters are incorporated, and $S$ = 12 mannually pre-defined orientations with an interval of 15 degrees are given. This setting will result in 24 filters by training only 2(each one with 12 different orientations).
Let the rotation matrices be $\{{\bf T}_s\}_{s=1}^{S}$, thus the parameterized filter for the $s$th orientation and the $j$th filter be ${\bf T}_s^T{\bf W}_j$, where ${\bf W}_j$ is the $j$th column of ${\bf W}$. Following \cite{sohn2012learning}, the energy function can be updated as:
\cutequationup
\begin{equation}
    E({\bf v},{\bf h};\boldsymbol{\theta})=-\sum_{j=1}^K\sum_{s=1}^S({\bf T}_s{\bf v})^T{\bf W}_jh_{j,s}-\sum_{j=1}^K\sum_{s=1}^Sb_{j,s}h_{j,s}-{\bf c}^T {\bf v}.
\cutequationdown
\end{equation}
Note that among all $S$ hidden nodes for the $j$th filter, at most one can be activated, i.e., $\sum_{s=1}^S h_{j,s}\leq1, \text{for } h_{j,s}\in\{0,1\}, j = 1,...,K$. This is achieved by considering $h_{j,s}$ an outcome of a multi-class classifier, with
\cutequationup
\begin{equation}
    \text{Pr}(h_{j,s}=1|{\bf v}) = \frac{\exp\left(({\bf T}_s{\bf v})^T{\bf W}_j+b_{j,s}\right)}{1+\sum_{s'=1}^S\exp(({\bf T}_{s'}{\bf v})^T{\bf W}_j+b_{j,s'})}.
\cutequationdown
\end{equation}
The conditional probability of the visible nodes will be kept the same as before. With these updates, training of the rotation-invariant filters follows the discussion in Sec.~\ref{sec:rbm}.

\cutsubsectionup
\subsection{Explanation of the network model}
\label{sec:vis}
The choice of the network configuration requires some explanation. The first four layers are used to extract features at increasing length scales. The 5th layer is added to further reduce the dimensionality of the data by learning the correlations among the 4th layer activations, i.e., the co-existence of global features in the samples. The number of nodes at layer 4th and 5th are manually configured to achieve a balance between high dimension reduction and low reconstruction error. However, we did not test all possibilities to confirm that our network design yields the best result.

\begin{figure}[htp]
\centering
\graphicspath{ {image/} }
\includegraphics[width=100mm,scale=0.5]{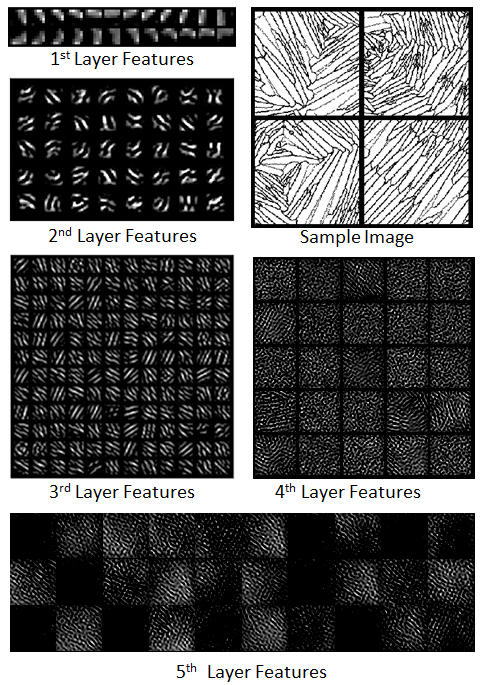}  
\caption{\textbf{Input Images and Extracted Features. Visualizations of filters for the 5 layers: The $1$st layer has two filters, each with 12 orientations (see Sec.~\ref{sec:orientation}). Due to limited space, Only the first 144 and 25 filters from the $3$rd and $4$th layers, respectively, are shown.}}
\label{fig:filter}
\end{figure}

\begin{figure}[H]
\centering
\graphicspath{ {image/} }
\includegraphics[width=100mm,scale=0.5]{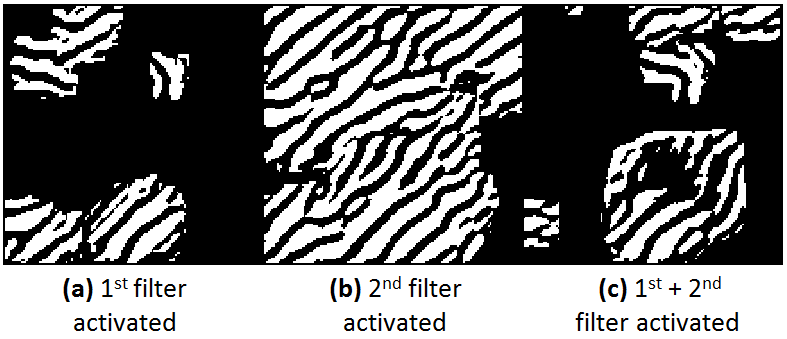}
\caption{\textbf{Activation on Different Filters. From left to right, reconstructions when only the $1^{st}$, only the $2^{nd}$, and both nodes from the 5th layer are activated, respectively}}
\label{fig:filter_activation}
\end{figure}

Fig.~\ref{fig:filter} visualizes filters of the five layers, learned from 100 Ti-6Al-4V alloy samples that are produced from the same laser sintering process. The visualization of each filter is done by iteratively convolving the filter (the network weights) with filters from the previous layers\footnote{This procedure is different from microstructure reconstruction (see Sec.~\ref{sec:post}) in that the sigmoid function is not applied to the convolution results.}. It is worth noting that the visualizations from the 5th layer include images of almost all black pixels. This is because the output images from these filters have almost uniform pixel values, appearing to be zeros after normalization. Also note that these filters are non-trivial, as the effect of activating multiple nodes at the output layer has nonlinear effect in the input space, i.e., the reconstructed image is not a simple addition of filter visualizations, because of the sigmoid operations throughout the network. To demonstrate, we show in Fig.~\ref{fig:filter_activation}(a)-(c) the reconstructions when the 1st, the 2nd, and both filters are activated.

\begin{figure}[htp]
\centering
\graphicspath{ {image/} }
\includegraphics[width=100mm,scale=0.5]{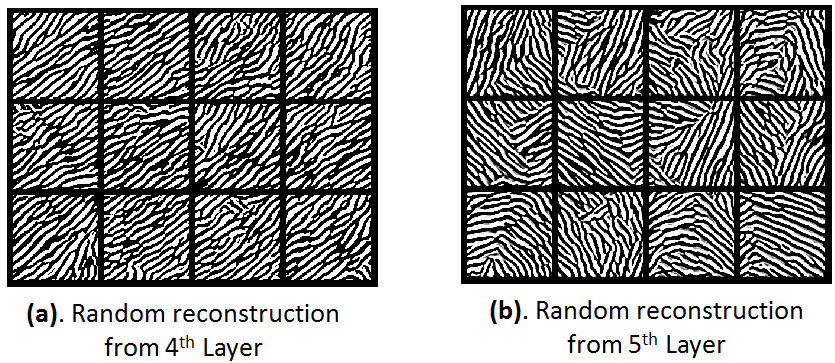}
\caption{\textbf{Comparison between random reconstructions from (a) the 4th and (b) the 5th layers, sampled from the corresponding design spaces ($\{0,1\}^{1000}$ in (a) and $\{0,1\}^{30}$ in (b)).}}
\label{fig:no5th}
\end{figure}

To further justify the network architecture, we observed through experiments that the random microstructure reconstructions obtained using the 5-layer network have larger variance than those from without the 5th layer, as illustrated in Fig.~\ref{fig:no5th}. This result indicates that the 5th layer is effective for avoiding the creation of repetitive microstructure samples.

\cutsubsectionup
\subsection{Reconstruction and postprocessing}
\label{sec:post}
\begin{figure}[htp]
\centering
\graphicspath{ {image/} }
\includegraphics[width=120mm,scale=0.7]{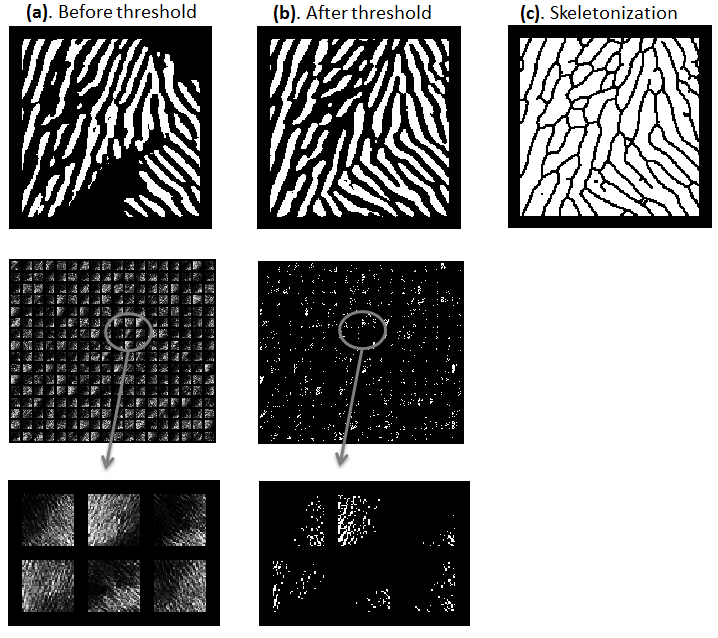}
\caption{\textbf{Postprocessing steps: Column (a) shows the original reconstruction and its  3rd layer activations (288 channels) with enlarged sample channels; Column (b) shows the reconstruction after thresholding the 3rd layer activations at 0.5; Column (c) shows the further improved reconstruction after skeletonization.}}
\label{fig:hidstate2}
\end{figure}
To reconstruct random microstructures, one starts by randomly assigning binary values to the last layer of the trained CDBN, and inversely sample the previous layers through deconvolution. To speed up the process, we directly treat activations (real values between 0 and 1) as the states of hidden nodes, to avoid drawing from Bernoulli distributions and eliminate randomness during the reconstruction. We discuss below the necessity of post-processing steps to avoid undesirable reconstructions. 

\begin{figure}[htp]
\centering
\graphicspath{ {image/} }
\includegraphics[width=120mm,scale=0.5]{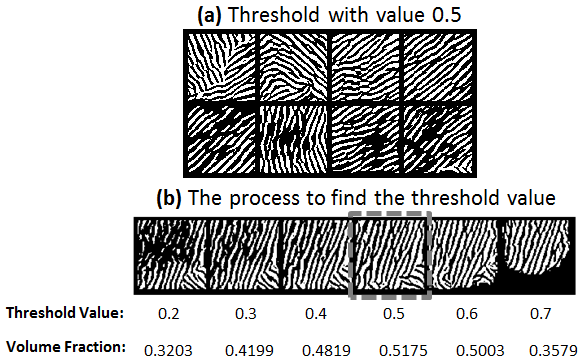}
\caption{\textbf{(a) A threshold of 0.5 does not always guarantee valid reconstructions, see the second row (b) Threshold can be fine-tuned based on a heuristic criterion: Here we test a set of threshold values and pick the one that yields a volume fraction closest to the average volume fraction of reconstructions of the original samples ($\rho=0.51$).}}
\label{fig:threshold05}
\end{figure}

First, we observe that with the proposed reconstruction method, the 3rd hidden layer (a $36\times36\times288$ image) is overly saturated, see Fig.~\ref{fig:hidstate2}a. It happens because the forward path of the proposed network has a Bernoulli sampling process, making the activation hidden layer binaries. But during the reconstruction, the sampling process is avoided in order to eliminate randomness. This causes the issue that the slightly activated nodes in the 3rd layer across its 288 channels, which should have a high chance to be sampled as zeros, to accumulatively affect the input layer, by creating overlapped grain boundaries (as black pixels) and undesirably rendering such regions as voids. Through experiments, we found that thresholding the activations of the 3rd layer at $\tau=0.5$ achieves the lowest average reconstruction error from the original Ti64 samples, see Fig.~\ref{fig:hidstate2}b. Some random reconstructions using this threshold are shown in Fig.~\ref{fig:threshold05}a. It is worth noting that this fixed threshold value may result in invalid reconstructions for some binary combinations at the output layer, see Fig.~\ref{fig:threshold05}a, two rows of the reconstruction microstructures are processed with the same threshold value $0.5$, but the performance is very different. The four images in the second row have more unexpected voids than the images in the first row. One potential solution is to fine-tune the threshold for each individual reconstruction based on a heuristic criterion, e.g., one can remove invalid voids in Ti64 reconstructions by thesholding the hidden layer so that the volume fraction of the resultant microstructure falls into a range calculated by valid reconstructions of the original samples, see Fig.~\ref{fig:threshold05}b for an example. Nonetheless, we are yet to identify thresholding criteria that are universal across all material systems.

Secondly, we notice that the resultant grain boundaries are wider than those from the original samples. This is due to the convolution and the max-pooling operations in the first three layers of the proposed CDBN. Applying an existing skeletonization method~\cite{skeleton} to the reconstruction results in the final outcome in Fig.~\ref{fig:hidstate2}c. Lastly, the convolution operation tends to smear the reconstructed image towards its boundary. The reported reconstruction results in this paper are the cropped areas ($143\times143$) without the blurred margins, which are resultant from the convolution.   

\cutsubsectionup
\subsection{Random reconstruction results}
\label{sec:recon}
\begin{figure}[htp]
\centering
\graphicspath{ {image/} }
\includegraphics[width=\linewidth]{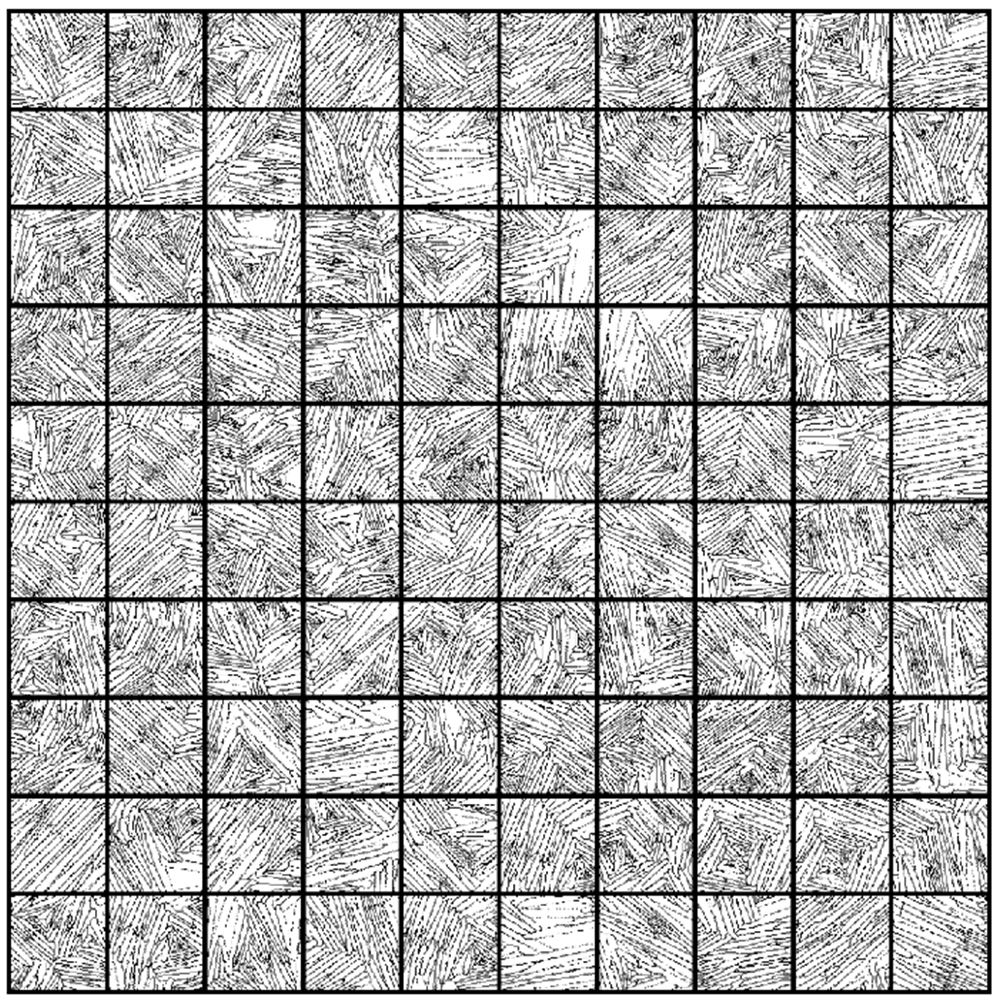}
\caption{\textbf{Ti-6Al-4V alloy}}
\label{fig:mat1}
\end{figure}

\begin{figure}[htp]
\centering
\graphicspath{ {image/} }
\includegraphics[width=0.9\linewidth]{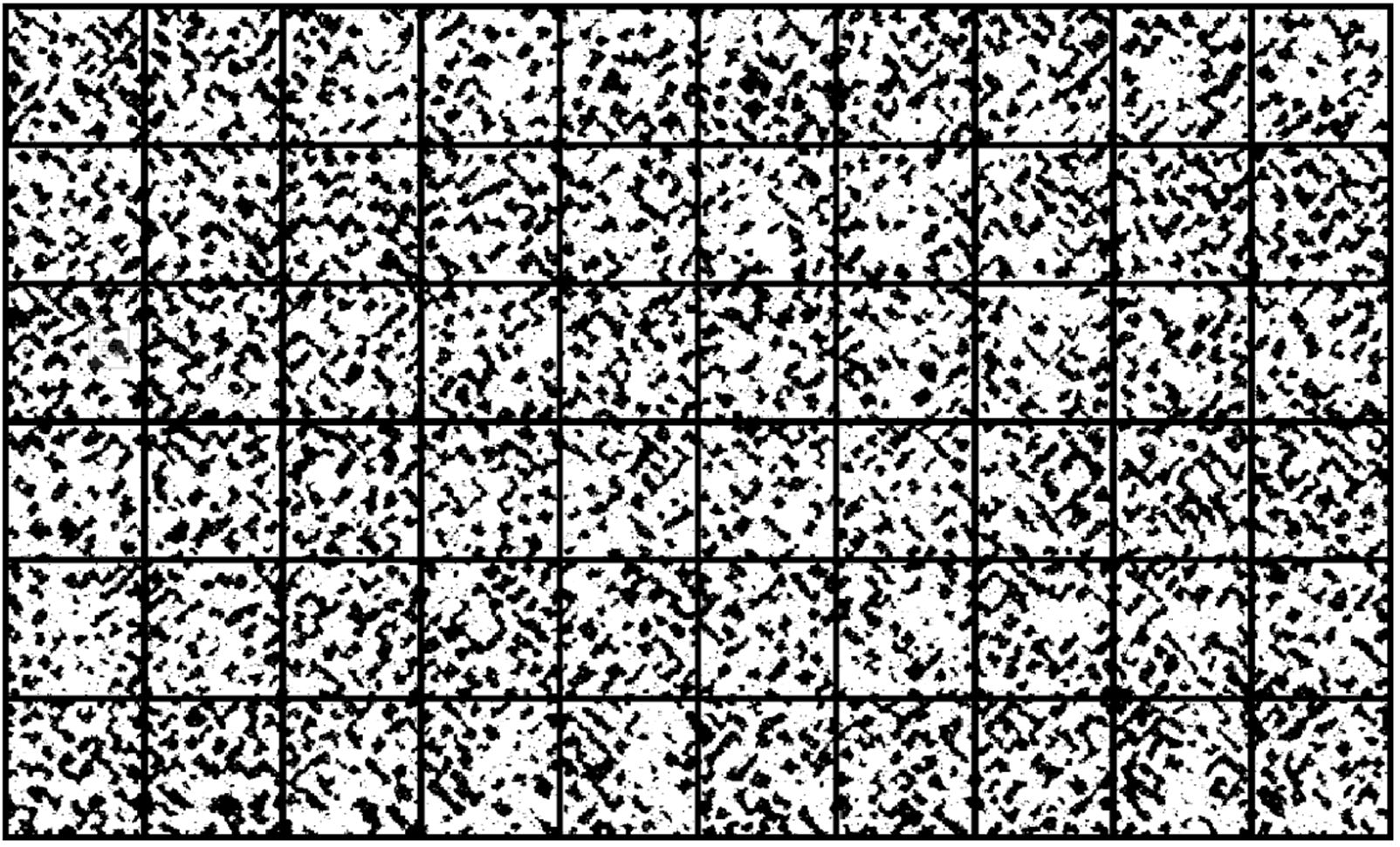}
\caption{\textbf{Pb-Sn (lead-tin) alloy}}
\label{fig:mat2}
\end{figure}

\begin{figure}[htp]
\centering
\graphicspath{ {image/} }
\includegraphics[width=0.9\linewidth]{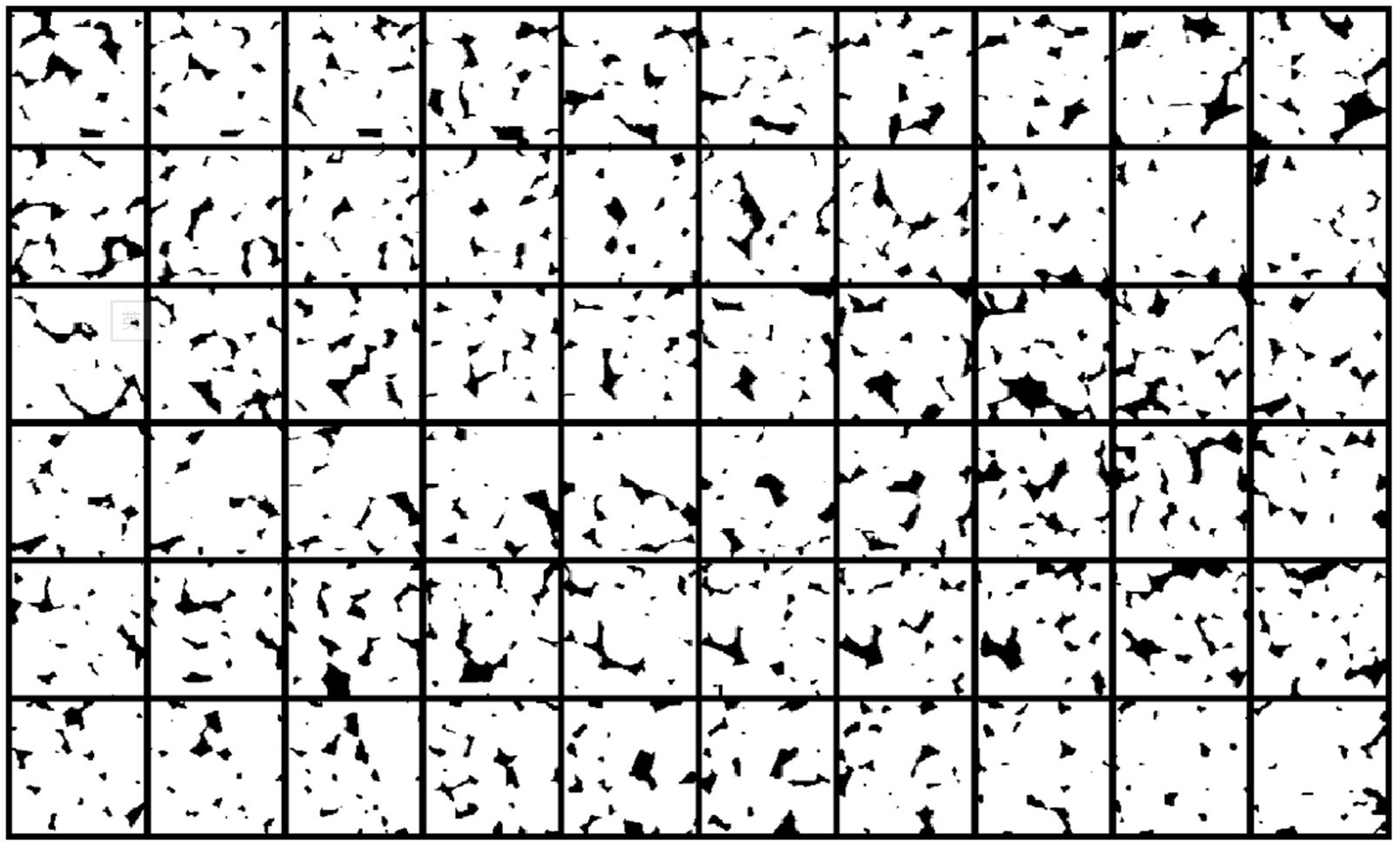}
\caption{\textbf{Pore structure of Fontainebleau sandstone}}
\label{fig:mat3}
\end{figure}

\begin{figure}
\centering
\graphicspath{ {image/} }
\includegraphics[width=0.9\linewidth]{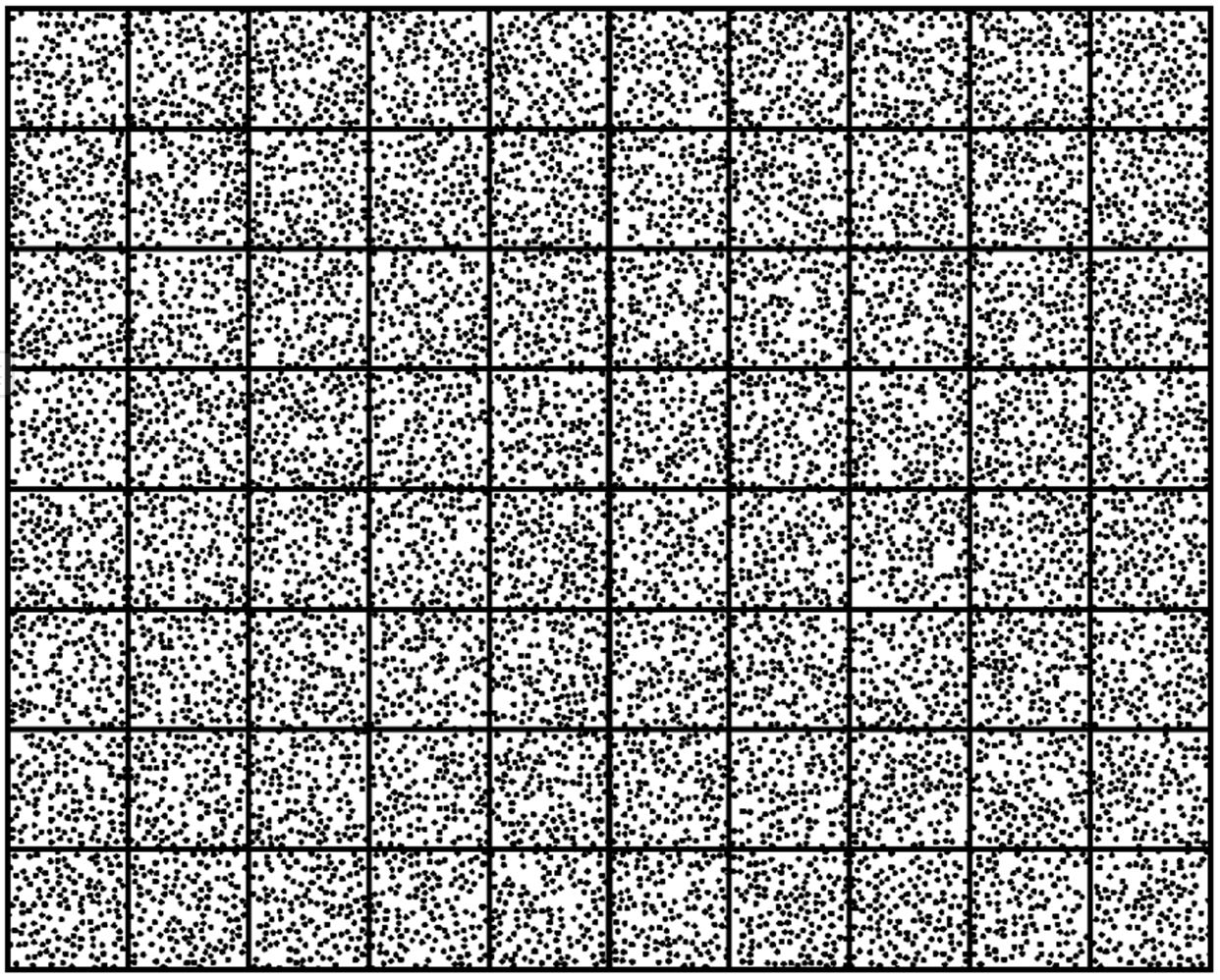}
\caption{\textbf{2D suspension of spherical colloids}}
\label{fig:mat4}
\end{figure}

Here we demonstrate the reconstruction performance of the proposed CDBN on four material systems: Ti-6Al-4V (100 images in Fig.~\ref{fig:mat1}), Pb-Sn(lead-tin) (60 images in Fig.~\ref{fig:mat2}), Pore structure of Fontainebleau sandstone (60 images in Fig.~\ref{fig:mat3}) and 2D suspension of spherical colloids (80 images in Fig.~\ref{fig:mat4}). Methods used to generate images of these four material systems are as follows: Ti64 are generated from physics-based kinetic Monte Carlo simulations~\cite{cheninpreparation} for beta to alpha phase transition starting from beta phase field; Pb-Sn optical images are generated from polished Pb-Sn solder ball surface~\cite{jiao2013modeling}; sandstone images are virtual 2D slices from 3D tomography reconstruction~\cite{li2016accurate}; and random sphere packing are generated using Monte Carlo simulations~\cite{li2014reconstruction}. 

\begin{figure}[htp]
\centering
\graphicspath{ {image/} }
\includegraphics[width=120mm,scale=0.7]{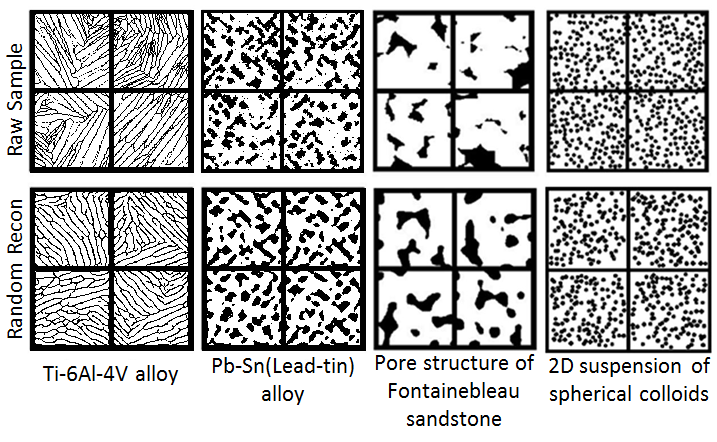}
\caption{\textbf{A comparison between random reconstructions (bottom) and the original samples (top) for four different material systems.}}
\label{fig:imagecompare}
\end{figure}

The CDBN specifications follow Sec.~\ref{sec:model}: Its input and output are a 200-by-200 image and a 30-dimensional binary vector, respectively. To generate microstructures, we can assign binary values to the 30 output neurons, and perform convolution backwards using the learned features. Fig.~\ref{fig:imagecompare} provides a visual comparison between samples and the corresponding random reconstructions. The first three material systems follow almost the same network setting, with the exception for Ti-6Al-4V alloy where the first layer is orientation-invariant. The network for spherical colloids is slightly different than the other three, since its universal local feature, a sphere, can be directly extracted from a single CRBM layer. Thus the rest four layers in this case are all fully-connected RBMs. From the results, it is evident that the proposed network has some general applicability across material systems. 

\cutsubsectionup
\subsection{Deeper understanding of the proposed model}
\label{sec:validation}
To further understand the performance and limitations of the proposed CDBN, we conduct two additional studies: The first concerns the restoration of original samples, and the second validates the reconstruction of the material property of interest, i.e., the critical fracture force computed using a recently developed volume-compensated lattice-particle (VCLP) method~\cite{chen2014generalized}.

\begin{figure}[h]
\centering
\graphicspath{ {image/} }
\includegraphics[width=\linewidth]{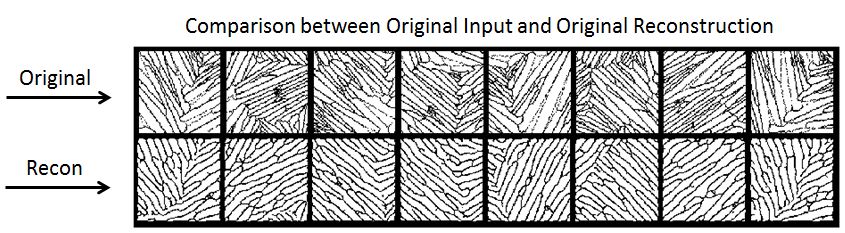}
\caption{\textbf{Comparison between the samples (top) and their reconstructions (bottom). Details of the two fail to match.}}
\label{fig:imagecompare3}
\end{figure}

For the first study, samples of the original Ti-6Al-4V alloy microstructure images and their corresponding reconstructions are shown in Fig.~\ref{fig:imagecompare3}. A close examination reveals discrepancies between the reconstructed and the original images. We believe that such discrepancies are caused by the nature of the network: Similar local patterns are learned as a single filter, and are replaced by this filter during reconstruction. Increasing the number of filters to learn may alleviate this issue, yet will also increase the computational cost for learning the network. Therefore a compromise is necessary. For example, for Ti64, Pb-Sn and Sandstone, we use 2 filters with 12 orientations in the first CRBM layer, and may fail to capture features (e.g., line segments) that do not belong to these pre-set orientations. The use of probabilistic max-pooling may also contribute to the discrepancies as forward and backward pooling do not preserve the activations. Note that preservation of local microstructure patterns can be achieved relatively easily for simple systems such as spherical colloids.

\begin{figure}[h]
\centering
\graphicspath{ {image/} }
\includegraphics[width=\linewidth]{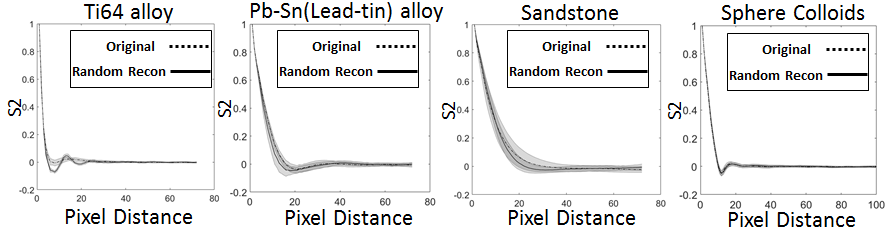}
\caption{\textbf{The 2-point correlation functions for the four different materials.}}
\label{fig:point_correlation}
\end{figure}

To quantitatively access the accuracy of the reconstructions, we computed the two-point correlation function $S_2(r)$, which gives the probability of two randomly selected points separated by distance $r$ falling into the same phase of interest, for both the samples and the associated reconstructions. Fig.~\ref{fig:point_correlation} shows the comparison. It can be seen that $S_2$ for the reconstructed microstructures statistically match the corresponding target structures well, except for the Ti64 system. We note that for the Ti64 system, both $S_2$ for the target and reconstruction exhibit clear oscillations for small $r$ values, indicating significant short range correlations due to the mutual exclusion volume between the grains. The wavelengths associated with the oscillations indicate the average grain width, and are the same in both functions. The oscillation in the reconstructed $S_2$ is apparently stronger, which is due to more uniform grain size and shape distribution in the reconstructed microstructure, and the missing of local details from the samples.

\begin{figure}[htp]
\centering
\graphicspath{ {image/} }
\includegraphics[width=\linewidth]{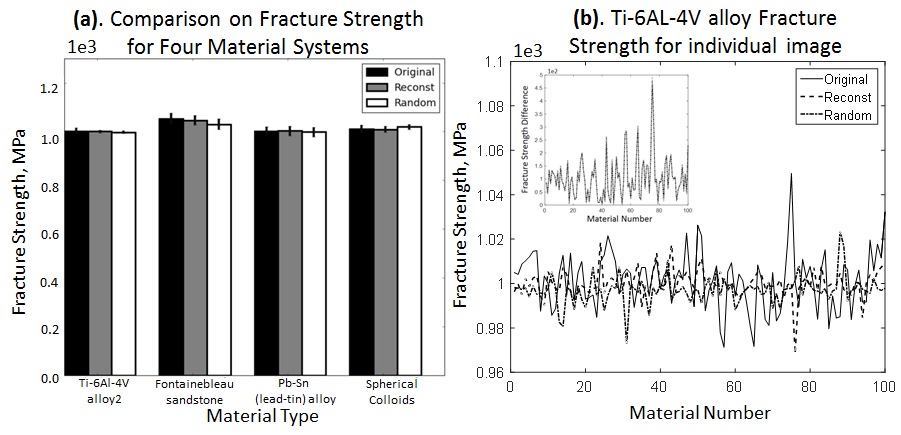}
\caption{\textbf{a) Comparison on critical fracture strength among the four material systems between the original samples (black), their corresponding reconstructions(gray) and random reconstructions (white) (b) Comparison on critical fracture strength among individual original images and the related reconstruction images (color online)}}
\label{fig:Simulation}
\end{figure}

In addition, we investigate the performance-wise discrepancies introduced by the reconstructions.
Specifically, we calculate the critical fracture force of three sets of samples: the original samples, their reconstructions, and 100 random reconstructions, using VCLP. The simulation results from four material systems are summarized in Fig.~\ref{fig:Simulation}, with two findings: (1) the average fracture forces of the three groups are statistically similar within each material system; and (2) discrepancies in the fracture force are visible between the original samples and their reconstructions, due to the aforementioned reconstruction error.
We argue that finding (1) is a desirable result, while individual-wise property matching may not be necessary. This is because each processing setting may produce various structures with small variance in their property values. Thus, during the structural design, it is required that the set of reconstructions of all samples derived from the same processing parameters (as is the case in this study) have statistically the same property as the original samples. 
\cutsectionup
\section{Discussion}
\label{sec:discussion}
Here we provide more discussions on the limitations and future directions of the presented work.

\cutsubsectionup
\subsection{Limitations}
First, we note that the network is designed to only reconstruct images of a fixed size, which is not desirable when scalable synthesis of material systems is needed. One potential solution is to apply a conditional probability model (e.g., \cite{bostanabad2016stochastic}) to the outputs of a hidden layer of the CDBN that represent the activations of key microstructure patterns. The intuition is that while the Markovian assumption may not directly hold at the image level (i.e., image pixels cannot be homogeneously inferred from their surroundings), the activations of local patterns may still be captured by this assumption. Such a hybrid model could be used to produce new microstructure patches given its surrounding patches, thus achieving synthesis of complex material samples. 

Secondly, identification of proper network specifications (e.g., the number of layers, the number of filters, and filter sizes) is non-trivial. In fact, an arbitrary choice of these parameters could lead to abysmal feature extraction and reconstruction performance. Due to the unknown sensitivity of the performance with respect to the modeling parameters, the authors cannot claim that the presented model is universally applicable to all material systems, despite the fact that the model worked well on the four demonstrated systems. 

\begin{figure}[htp]
\centering
\graphicspath{ {image/} }
\includegraphics[width=100mm,scale=0.5]{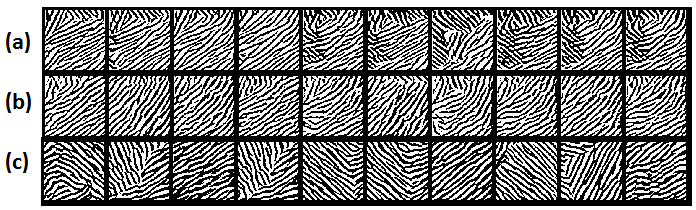}
\caption{\textbf{(a-c) Sample reconstructions (without skeletonization) based on models trained from 10, 50, and 100 samples, respectively. The variances among random reconstructions for these three cases are 0.2644, 0.2863, and 0.4205, respectively. The variance is calculated from 100 reconstructions in each case.}}
\label{fig:different_training_samples}
\end{figure}

Thirdly, the size of training samples can significantly affect the reconstruction performance. To show this, we learned three network models based on 10 and 50 training samples randomly drawn from the 100 Ti64 images. We then reconstruct 100 random images using each of the trained models and image-wise variances for each model are calculated as 0.2644, 0.2863, and 0.4205 respectively. As expected, the result shows that a larger training data leads to more distinct filters, which leads to the reconstructions with high variance Fig.~\ref{fig:different_training_samples}.



Lastly, it is also worth noting that the proposed network essentially learns a lower-dimensional manifold from the image space using limited samples. However, it does not provide a feasible region within this manifold. As a result, random reconstructions are not guaranteed to be physically meaningful, and a validation through processing-structure mapping is needed. It is therefore desirable that certain known physics-based rules (i.e., constraints on features) can be incorporated into the feature extraction process to eliminate or reduce the chance of creating infeasible reconstructions. 

\cutsubsectionup
\subsection{From material reconstruction to material design}
As mentioned in the introduction, the feature learning method proposed in this paper has two potential contributions to computationally efficient material design: The reduced design space will enable more tractable search for complex microsturcture designs, and allows statistical models with better prediction performance to be built for process-structure-property mappings. Achieving the latter, however, requires the learned features to be able to explain variance in material properties (in the structure-property mapping), and in processing settings (in the process-structure mapping). While this study does not directly demonstrate the connection between features and properties or processing parameters, successful reconstructions through the developed model showed its potential at learning microstructure patterns, which could be fine-tuned through a supervised model to statistically explain properties or processing parameters.

\cutsectionup
\section{Conclusions}
\label{sec:con}
This paper presented a novel methodology for feature extraction and reconstruction of complex microstructures through a convolutional deep belief network. 
Using four different material systems, we showed that the proposed 5-layer CDBN model, along with postprocessing techniques, can achieve significant dimension reduction,visually and statistically plausible random reconstruction, and statistically preserve the critical fracture strength values after reconstruction. Key limitations include (1) the lack of scalability in reconstruction, (2) the individual discrepancies in fracture strength between original and reconstructed images, and (3) the lack of guarantee of the validity of the reconstructions, i.e., the threshold for the reconstruction is only empirically chosen. Nonetheless, the demonstrated advantages over existing material reconstruction approaches showed that our method could lead to more efficient representations of complex microstructures, which is critical to accelerating ICME.


\cutsectionup
\section*{Acknowledgement}
\label{sec:ack}
This work is partially supported by 
NSF CMMI under grant No.
1651147 (Program Manager: Mary Toney) and by 
DOD DARPA under grant No.
N66001-14-1-4036 (Program Manager: Fariba Fahroo, DARPA Mentor:
Michael Maher). R. C. and Y. R. thank the startup funding from the
Arizona State University. The authors thank Dr. Honglak Lee, Dr.
Kihyuk Sohn and Ye Liu for their support on the CDBN
implementation, and Dr. Wei Chen, Dr. Hongyi Xu, Dr. Ramin Bostanabad
for valuable discussion. All source codes and datasets are
available at
\url{https://github.com/DesignInformaticsLab/Material-Design}.


\bibliographystyle{elsarticle-num}
\bibliography{reference2.bib}

\cutsubsectionup

\cutsubsectionup
\section*{Appendix: Sample Microstructure Images}
\label{sec:appendixB}

\clearpage
\thispagestyle{empty}

\listoftables
\listoffigures

\end{document}